\documentclass[aps,prb,10pt,balancelastpage,letterpaper,superscriptaddress,longbibliography,twocolumn]{revtex4-1}

\usepackage{times}

\usepackage{amsmath}
\usepackage{amssymb}
\usepackage{graphicx}
\usepackage{bm}
\usepackage{bbm}

\usepackage{color}
\usepackage{verbatim}
\usepackage{multirow}
\usepackage{hyperref}

\newcommand  {\suppcaption}[1] {\parbox{\textwidth}{#1}}

\newcommand  {\avg}[1]       {\langle#1\rangle}

\newcommand  {\ii}           {\mathrm{i}}

\newcommand  {\bramidket}[3] {\langle#1\vert#2\vert#3\rangle}

\newcommand  {\spinup}       {\mathord{\uparrow}}
\newcommand  {\spindn}       {\mathord{\downarrow}}
\newcommand  {\eV}           {\,\mathrm{eV}}
\newcommand  {\meV}          {\,\mathrm{meV}}
\newcommand  {\nm}           {\,\mathrm{nm}}

\newcommand  {\nn}[1]        {\left\langle #1 \right\rangle}
\newcommand  {\nnn}[1]       {\left\langle\langle #1 \right\rangle\rangle}
\renewcommand{\vec}[1]       {\mathbf{#1}}                        
\renewcommand{\vr}           {\vec{r}}
\newcommand  {\vecsigma}     {\bm{\sigma}}

\begin{document}

\title{Topological states in multi-orbital HgTe honeycomb lattices }

\author{W. Beugeling}
\affiliation{Max-Planck-Institut f\"ur Physik komplexer Systeme, N\"othnitzer Stra\ss e 38, 01187 Dresden, Germany}

\author{E. Kalesaki}
\affiliation{IEMN- Dept. ISEN, UMR CNRS 8520, 41 boulevard Vauban, 59046 Lille Cedex, France}
\affiliation{Physics and Materials Science Research Unit, University of Luxembourg, 162a avenue de la Fa\"iencerie, L-1511 Luxembourg}
\author{C. Delerue}
\affiliation{IEMN- Dept. ISEN, UMR CNRS 8520, 41 boulevard Vauban, 59046 Lille Cedex, France}

\author{Y.-M. Niquet}
\affiliation{Univ. Grenoble Alpes, INAC-SP2M, L\_Sim, Grenoble, France and CEA, INAC-SP2M, L\_Sim, 17 avenue des Martyrs, 38054 Grenoble, France}

\author{D. Vanmaekelbergh}
\affiliation{Debye Institute for Nanomaterials Science, Utrecht University, Princetonplein 1, 3584 CC Utrecht, The Netherlands}

\author{C. Morais Smith}
\affiliation{Institute for Theoretical Physics, Center for Extreme Matter and Emergent Phenomena, Utrecht University, Leuvenlaan 4, 3584 CE Utrecht, The Netherlands}

\date{\today}

\begin{abstract}
{\bf Research on graphene has revealed remarkable phenomena arising in the honeycomb lattice.
However, the quantum spin Hall effect predicted at the $K$ point could not be observed in
graphene and other honeycomb structures of light elements due to an insufficiently strong
spin-orbit coupling. Here we show theoretically that 2D honeycomb lattices of HgTe can
combine the effects of the honeycomb geometry and strong spin-orbit coupling. The
conduction bands, experimentally accessible via doping, can be described by a tight-binding
lattice model as in graphene, but including multi-orbital degrees of freedom and spin-orbit
coupling. This results in very large topological gaps (up to $35\meV$) and a flattened band
detached from the others. Owing to this flat band and the sizable Coulomb interaction,
honeycomb structures of HgTe constitute a promising platform for the observation of a
fractional Chern insulator or a fractional quantum spin Hall phase.
}
\end{abstract}

\maketitle

\onecolumngrid
{\footnotesize Correspondence and requests for materials should be addressed to C.D. (email: christophe.delerue@isen.fr) or to C.M.S. (email: C.deMoraisSmith@uu.nl).}

\hfill \break
\twocolumngrid

The discovery of graphene has confronted us with a material that exhibits fascinating electronic properties \cite{CastroNetoEA2009}, such as zero-mass carriers, persisting conductivity at vanishing density at the Dirac point \cite{Nair08}, Klein tunnelling \cite{Katsnelson06}, and an anomalous quantum Hall effect \cite{NovoselovEA2005,DuEA2009,BolotinEA2009}. Nevertheless, the absence of a band gap in its spectrum prevents its use as a field-effect transistor, and its weak spin-orbit coupling (SOC) hampers the possibility to realize the quantum spin Hall effect (QSHE) \cite{KaneMele2005PRL95-22} and use it for quantum spintronics. The prospect of artificial graphene  samples \cite{Polini13} that display the lacking properties has motivated research in various types of honeycomb lattice, such as arrays of ultracold atoms \cite{SoltanPanahiEA2011}, molecular graphene \cite{GomesEA2012}, organometallic lattices \cite{Wang13}, and 2D electron gases subject to a geometric array of gates \cite{Gibertini09,ParkLouie2009}. 
Recently, an alternative path came from the self-assembly of semiconductor nanocrystals forming atomically coherent 2D structures with a long-range honeycomb pattern; the thickness and honeycomb period are defined by the size of the nanocrystals, which is in the range of 5 nm \cite{EversEA2013,BoneschanscherEA2014}.
Honeycomb lattices of PbSe and CdSe nanocrystals have been fabricated, with astonishing atomic coherence due to the oriented attachment of the nanocrystals. Theoretical investigations have shown that CdSe superlattices formed in such a way exhibit Dirac cones at two energies and nearly dispersionless bands. Unfortunately, these flat bands are connected to the nearby higher energy bands, and the SOC gaps in the conduction band are very small \cite{KalesakiEA2014}.

In this work, we propose a design for robust topological
insulators that combine the properties of the honeycomb lattice
and strong SOC. We consider three different types of HgTe layers
with superimposed honeycomb geometry and present atomistic
tight-binding (TB) calculations of their conduction band
structure that we accurately describe by a 16-band effective
model. Such lattices take advantage of the multi-orbital degrees of
freedom in the honeycomb setup, allied to the strong SOC\cite{ZhangEA2014}. The access to multi-orbital degrees of freedom allows for further manipulation of the topological properties. 
While the Haldane \cite{Haldane1988} and Kane-Mele \cite{KaneMele2005PRL95-22} models originally concerned honeycomb lattices characterized by a single orbital per site (e.g., $p_z$ in graphene) and isotropic nearest-neighbour (NN) hopping integrals between them, multi-orbital models have attracted much attention recently, in particular because they seem to be a paradigm to generate topologically non-trivial flat bands \cite{WuEA2007,SunEA2011,Venderbos11,Yang12,Olschlager13}. The flat band structure opens the way to the realization of interesting strongly correlated states, such as fractional QSHE, fractional Chern insulators, or ferromagnetic fractional Chern insulators \cite{NeupertEA2011,SunEA2011,HuEA2011,Goerbig2012,RegnaultBernevig2011,TangEA2011,Xiao2011,Yang12,HeEA2012,Bergholtz13}.
The currently proposed HgTe lattices exhibit
conduction bands characterized by large topological gaps and an
isolated flat band. We conclude that, depending on the position of
the Fermi level, not only QSHE could be observed in these
structures, but also fractional QSHE or fractional Chern insulator
phases, as on-site and NN Coulomb-interaction parameters are
found in the energy range required for their realization.

\hfill \break

{\bf Results}


\begin{figure*}[t]
\centering
\includegraphics[width=\textwidth]{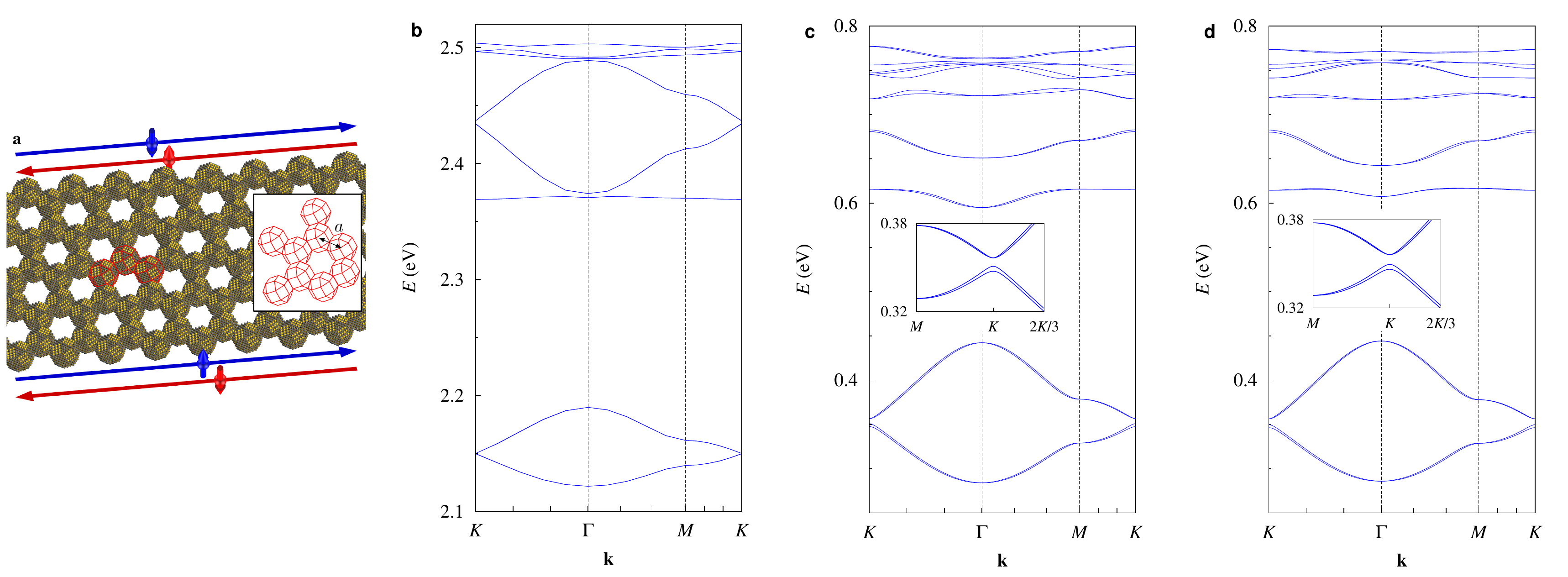}
\caption{\textbf{Nanocrystal lattices and their conduction band dispersions.} \textbf{a}, honeycomb nanoribbon formed by the HgTe (CdSe) nanocrystals. Hg (Cd) atoms are in yellow, Te (Se) atoms are in grey. Each nanocrystal has a truncated nanocube shape, the vertices of which are given by all permutations of $[\pm 1, \pm (1-q), \pm (1-q)]l$, where $q$ is the truncation factor and $2 l$ is the size of the original nanocube before truncation. The honeycomb lattice spacing $a$, i.e., the center-to-center distance between neighbor nanocrystals, is defined as $a=(2N+1)a_{0}/\sqrt{2}$ where $N$ is an integer and $a_0$ is the cubic lattice parameter of HgTe (CdSe). The nanocrystals are attached via $\langle 110 \rangle$ facets ($a=\sqrt{2}(2-q)l$). The arrows along the ribbon indicate the electron propagation in the helical edge states present in the quantum spin Hall phase. Red and blue colors correspond to top and bottom edge for spin up, bottom and top edge for spin down, respectively. \textbf{b,c}, band dispersions for the bulk resulting from the atomistic tight-binding calculations with $q=0.5$ ($a=5.0\nm$ for HgTe, $4.7\nm$ for CdSe). \textbf{d}, same for the HgTe superlattice resulting from the effective model. Insets in (\textbf{c,d}) show the $s$ bands in the gap region with higher magnification $\left[ 2K/3 = \left(\frac{4 \pi}{9a} , \frac{4 \pi}{9 \sqrt{3} a} \right) \right]$.}
 \label{fig_bulkdispTB}
\end{figure*}


{\bf Design of the HgTe honeycomb lattices.} The three structures that we consider have in common that the $[111]$ direction of the zincblende lattice is perpendicular to the plane. The first system is inspired by recent results obtained for nanocrystal self-assembly \cite{EversEA2013,BoneschanscherEA2014}, i.e., a graphene-type superlattice of truncated cubic nanocrystals attached via $\langle110 \rangle $ facets. The second system consists of spheres connected by short cylinders, which allow us to vary the coupling between honeycomb lattice sites in a convenient way. The third system corresponds to a honeycomb array of cylinders. These two last systems may be experimentally realized by gas-phase deposition and lithography. We will show below that the multi-orbital topological effects are common to these three different structures.

The HgTe superlattices proposed here differ fundamentally from HgTe/CdTe quantum wells, where the QSHE has been predicted \cite{BernevigEA2006} and experimentally observed \cite{KonigEA2007}. In the latter system, the appearance of the QSHE at the $\Gamma$ point is connected to band inversion and vanishes for quantum wells of thickness below 6 nm. Here, instead, the effect occurs at the $K$ point, and it is driven by the honeycomb nanogeometry, allied to the strong SOC of the composing HgTe nanocrystals. This distinction is important because zero modes (Majoranas) bind to topological lattice defects when the band gap opens at a non-$\Gamma$ point in the Brillouin zone  \cite{RanEA2009,JuricicEA2012}, and hence, the underlying topological order can be detected by measuring the structure of the topological defect \cite{RueggEA2013,SlagerEA2013}.

{\bf Band structure of lattices of HgTe nanocrystals.} To unveil the topological properties of these systems, we have performed atomistic tight-binding (TB) band-structure calculations. We use a basis of twenty atomic orbitals on each atom of the nanocrystals, including the spin degree of freedom. The methodology is described in Ref.~\onlinecite{Allan12} and is summarized in the Methods section. In comparison to bulk HgTe, the electronic structure of HgTe superlattices is characterized by a large bandgap due to the strong quantum confinement. Conduction and valence bands close to this gap are composed of several minibands and minigaps due to the periodic scattering of the electronic waves in the honeycomb structure \cite{KalesakiEA2014}. In the following, we only discuss the physics of the sixteen lowest conduction bands (Fig.~\ref{fig_bulkdispTB}).

A typical honeycomb lattice composed of HgTe nanocrystals is shown in Fig.~\ref{fig_bulkdispTB}a and the related conduction-band structure is displayed in Fig.~\ref{fig_bulkdispTB}c. In order to visualize the effects of SOC, we compare it to the band structure of the same honeycomb structure composed of CdSe (Fig.~\ref{fig_bulkdispTB}b). The strong SOC in HgTe gives rise to effects which are absent in CdSe \cite{KalesakiEA2014}. The overall behaviour of the band structure can be understood as follows. Each individual nanocrystal is characterized by two states with $s$ envelope wave-function and six $p$ states at higher energy. In the honeycomb structure, strong coupling between the wave-functions of neighbour nanocrystals leads to the formation of sixteen bands grouped into two manifolds of four ($s$) and twelve ($p$) bands that are well separated. The $s$ bands have the same type of dispersion as the $\pi$ and $\pi^{\star}$ bands in real graphene \cite{CastroNetoEA2009}. In the case of CdSe (Fig.~\ref{fig_bulkdispTB}b), these bands are spin-degenerate and are connected at the $K$ and $K'$ points of the Brillouin zone, where their dispersion is linear (Dirac points). In HgTe, instead, the $s$ bands exhibit a small gap ($5.7\meV$) at the $K$ points and  have a quadratic dispersion (see Fig.~\ref{fig_bulkdispTB}c). In addition, they are characterized by a visible spin splitting at all points of the Brillouin zone, except at $\Gamma$ and $M$. Among the twelve $p$ bands of CdSe nanocrystal superlattices, eight have a small dispersion and the other four basically behave like the (Dirac) $s$ bands. Four flat bands are built from the $p_z$ states perpendicular to the lattice, which are not very dispersive because $p_z$-$p_z$ ($\pi$) interactions are weak. Four other bands ($p_{x,y}$), respectively above and below the $p$ Dirac band, are flat due to destructive interferences of electron hopping induced by the honeycomb geometry \cite{WuEA2007,SunEA2011}. In honeycomb lattices of HgTe nanocrystals, the SOC induces spin splitting, opens a large gap at $K$ in the $p$-like Dirac bands, and produces a considerable detachment of the lowest flat $p$-band from the Dirac $p$-band (Fig.~\ref{fig_bulkdispTB}c). The effects of the SOC are so strong that it is hardly possible to recognize the Dirac bands.

These unexpected features allow for the realization of several topological states of matter, by doping the system using a field-effect transistor or electrolyte gel gating \cite{YeEA2010}. At zero energy, the undoped system is a semiconductor with a trivial gap of about 0.4 eV. Upon doping the material with one electron per nanocrystal, the small $s$-like topological gap may be reached, whereas for fillings between 2 and 3 electrons per nanocrystal, the fractional quantum (spin) Hall regime may be realized at the flat band. For a doping level of 4 electrons/nanocrystal one reaches the QSHE gap. At this point, we should emphasize that doping of nanocrystals with up to 10 electrons has already been demonstrated experimentally \cite{Ger2003}; therefore, all the interesting regimes that we discuss are at reach with the existing technology, at a simple switch of the doping level. Other examples of band structures for lattices of HgTe nanocrystals with different size or truncation factors are presented in Supplementary Fig.~1 and are discussed in Supplementary Note 1. They all show large topological gaps, especially in the $p$ sector.


\begin{figure*}[t]
\centering
\includegraphics[width=0.7\textwidth]{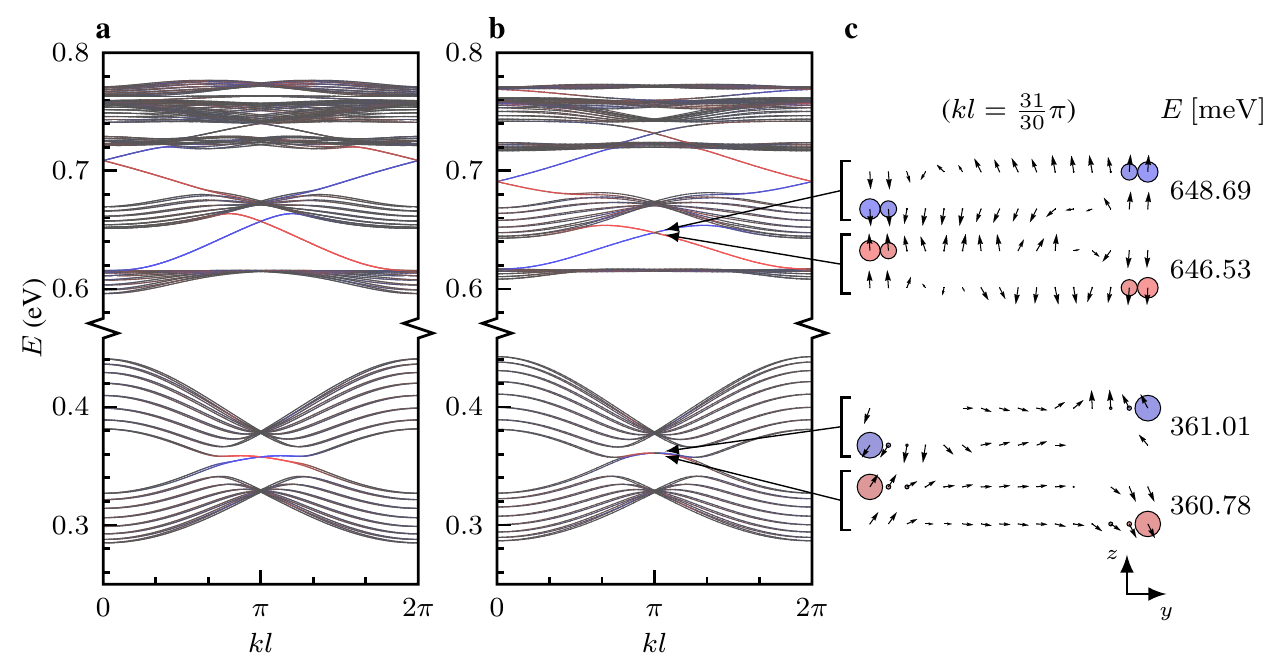}
\caption{\textbf{Topological edge states and non-trivial gaps in honeycomb lattices of HgTe nanocrystals.} \textbf{a}, conduction bands calculated using the atomistic tight-binding method for a zigzag ribbon composed of sixteen nanocrystals per unit cell ($q=0.5$, body diagonal of $5.0\nm$, cell length $l=8.7\nm$). \textbf{b}, same but computed from the effective Hamiltonian. \textbf{c}, spin orientation on each site for a selection of states calculated at $k = 31\pi /(30 l)$. A vertical arrow indicates that the spin is along the $z$ direction, perpendicular to the lattice. The size of the circles represents the weight of the wave function on each site. In each figure, the colour indicates the expectation value $\avg{y\sigma_z}$, i.e., red and blue correspond to top and bottom edge for spin up (bottom and top edge for spin down), respectively. At each energy $E$, there are two states which live on opposite edges with opposite spin, therefore they are represented by the same colour. The bulk states are grey. }
  \label{fig_edge_disp}
\end{figure*}


The topological properties of the bands are most transparently studied through an edge-state analysis in a 1D nanoribbon \cite{BeugelingEA2012PRB86-07}. We consider a zigzag ribbon composed of sixteen nanocrystals ($34740$ atoms) per unit cell. Figure~\ref{fig_edge_disp}a shows that edge states are crossing the three gaps between the $p$ bands as well as the gap between the $s$ bands. These results also hold for armchair ribbons. In Supplementary Fig.~2a, we present the band structure for another nanoribbon, which has two inequivalent edges. Still, helical edge states characteristic of the QSHE are found, as shown by the 2D plots of the wavefunctions (Supplementary Figs. 2b--e).

{\bf Effective model.} The band structures resulting from the atomistic TB calculations are accurately described by a sixteen-band effective model (Fig.~\ref{fig_bulkdispTB}d), where each nanocrystal is treated as one site on a honeycomb lattice. The effective TB model is written in the basis of the four aforementioned orbitals ($s, p_x,p_y,p_z$) per site as $ H_\mathrm{eff} = H_\mu + H_\mathrm{NN} + H_\mathrm{ISO} + H_\mathrm{RSO}$. Here, $H_\mu$ incorporates the on-site energies $E_s$, $E_{p_x}=E_{p_y}$, and $E_{p_z}$. The edge nanocrystals have a slightly different value of $E_s$ compared to the bulk, in order to account for the different number of neighbours. The term 
\begin{equation}\label{eqn_hnn}
  H_{\mathrm{NN}} = \sum_{\nn{i,j}} \sum_{\substack{\alpha\\b,b'}} c^\dagger_{i,b,\alpha}V_{i,b;j,b'}c_{j,b',\alpha},
\end{equation}
encodes NN hopping, where $\nn{i,j}$ denotes NN sites, $\alpha=\spinup,\spindn$ denotes spin, and $b$ and $b'$ the orbitals. The coupling coefficients $V_{i,b;j,b'}$ are expressed in terms of the hopping parameters $V_{ss\sigma}$, $V_{pp\sigma}$, $V_{pp\pi}$, and $V_{sp\sigma}$, following the notations of Ref.~\onlinecite{SlaterKoster1954}. 

The intrinsic SOC term $H_\mathrm{ISO}$ couples the electron orbital angular momentum $\vec{L}$ and spin $\vec{S}=\vecsigma/2$. In the $p$ sector, it is encoded through the on-site term $\lambda_\mathrm{ISO}^{p} \vec{L} \cdot \vecsigma$. There is no on-site term in the $s$ sector because the orbital angular momentum is ``frozen''. For the same reason, in graphene, the on-site intrinsic SOC term is absent because the $sp^2$ hybridization freezes the orbital momentum in the $p_z$ state. However, as shown by Kane and Mele \cite{KaneMele2005PRL95-22}, the intrinsic SOC introduces a next-nearest-neighbour (NNN) hopping term, which is written as
\begin{equation}
H_{\mathrm{ISO}} = \ii\lambda_\mathrm{ISO}^{s}\sum_{\nnn{i,j}}\sum_\alpha c^\dagger_{i,s,\alpha} \sigma^z_{\alpha\alpha}\nu_{ij}c_{j,s,\alpha}. 
\end{equation}
Here, the summation is over NNNs, and $\nu_{ij}=\pm 1$, with the sign depending on the outer product of the two NN vectors that connect sites $i$ and $j$. The Rashba SOC term, proportional to the cross product of momentum and spin, $\vec{p}\times \vec{S}$, is written as a NN-hopping term
\begin{equation}\label{eqn_rashba}
H_{\mathrm{RSO}} = \ii \sum_{\nn{i,j}} \sum_{\substack{\alpha,\beta\\b,b'}} c^\dagger_{i,b,\alpha}\gamma_{i,b;j,b'}[\hat{z}\cdot(\vecsigma\times \vr_{ij})]_{\alpha\beta}c_{j,b',\beta}. 
\end{equation}
The coupling coefficients $\gamma_{i,b;j,b'}$ have the same structure as the $V_{i,b;j,b'}$ for the ordinary NN hopping and are expressed in terms of $\gamma_{ss\sigma}$, $\gamma_{pp\sigma}$, and $\gamma_{pp\pi}$. The $sp$ term may be neglected due to the large energy separation between the $s$ and $p$ bands.


\begin{table*}
  \center%
  \begin{tabular}{llll}
    \multicolumn{4}{c}{}\\
    on-site             & NN hopping              & Rashba SOC                         & intrinsic SOC\\
    \hline
    $E_s^\mathrm{bulk} = 0.365\eV$ & $V_{ss\sigma} =-26.4\meV$ & $\gamma_{ss\sigma} = 0.56 \meV$ & $\lambda_\mathrm{ISO}^{s} = 0.71\meV$\\
    $E_s^\mathrm{edge} = 0.370\eV$ & $V_{pp\sigma} = 45.6\meV$ & $\gamma_{pp\sigma} = 1.50 \meV$ & $\lambda_\mathrm{ISO}^{p} = 15.8\meV$\\
    $E_{p_x} = 0.691\eV$           & $V_{pp\pi}    = -2.7\meV$ & $\gamma_{pp\pi}    = 0.80 \meV$ \\
    $E_{p_y} = 0.691\eV$           & $V_{sp\sigma} = 31.1\meV$ \\
    $E_{p_z} = 0.747\eV$

  \end{tabular}
  \caption{{\bf Parameters of the effective model.} Parameters derived for the lattice of HgTe nanocrystals described in Fig.~\ref{fig_bulkdispTB}. $E_s$, $E_{p_x}$, $E_{p_y}$, and $E_{p_z}$ are the on-site energies on the $s$, $p_x$, $p_y$, and $p_z$ orbitals, respectively. In ribbons, the edge nanocrystals have a slightly different value of $E_s$ compared to the bulk ($E_s^\mathrm{bulk} \neq E_s^\mathrm{edge}$). $V_{ss\sigma}$, $V_{pp\sigma}$, $V_{pp\pi}$, and $V_{sp\sigma}$ are the hopping parameters, following the notations of Ref.~\onlinecite{SlaterKoster1954}. $\gamma_{ss\sigma}$, $\gamma_{pp\sigma}$, and $\gamma_{pp\pi}$ are the terms describing the Rashba SOC, following the same notations. The intrinsic SOC is defined by $\lambda_\mathrm{ISO}^{s}$ and $\lambda_\mathrm{ISO}^{p}$ on $s$ and $p$ orbitals, respectively.}
  \label{tbl_parameters}
\end{table*}


In Table~\ref{tbl_parameters}, we present typical values for the parameters obtained numerically using least-squares fitting to the band structure of Fig.~\ref{fig_bulkdispTB}c. The band structure of the effective model is shown in Fig.~\ref{fig_bulkdispTB}d. As expected, the on-site term $\lambda_\mathrm{ISO}^{p}$ is much larger than the NNN term $\lambda_\mathrm{ISO}^{s}$, explaining the opening of a very large gap at the $K$ point and the detachment of the flat bands in the $p$ sector. Using the effective model, the non-trivial topology of the bands is confirmed by the calculation of the $Z_2$ topological invariant, the spin Chern number (Methods). 

The effective model yields a band structure for the ribbon (Fig.~\ref{fig_edge_disp}b) in excellent agreement with the atomistic TB calculations (Fig.~\ref{fig_edge_disp}a). The red and blue colours in Figs.~\ref{fig_edge_disp}a,b indicate the expectation value $\avg{y\sigma_z}=\bramidket{\psi_i}{\hat{y}\hat{\sigma}_z}{\psi_i}$, where $y$ is the coordinate perpendicular to the ribbon edges. This expectation value allows us to identify helical edge states that come in pairs with identical dispersion, opposite spin, and live on opposite edges. Both atomistic and effective TB models show that the gaps in the $s$ and $p$ sectors exhibit helical edge states, characteristic of the QSHE.

When the Rashba coupling is neglected in the effective model, all states are spin degenerate. The Rashba term induces a small splitting in energy, and tilts the spins slightly away from the perpendicular direction. In Figs.~\ref{fig_edge_disp}c, we plot the spin direction on each site of the zigzag ribbon for four sets of four edge states (two in each edge). The spin direction is always perpendicular to the edge, i.e., the spin lies in the $yz$ plane if we choose the $x$ direction to be parallel to the edge. The localization of the selected states on the edges is visible from the weight of the wave function, indicated by the size of the circles in the figure. The colours of the circles are determined by the local value of $y\sigma_z$, and correspond to the colours in the dispersion (Fig.~\ref{fig_edge_disp}a,b). The site dependence of the spin direction leads to interesting spin textures. For the edge states, the typical spin texture is almost smooth: Going from one edge to the other, the spin direction interpolates between (almost) up and (almost) down in a rotational manner. A slight tilt of $\sim 3^{\circ}$ is observed for the edge states in the $p$ bands. In the $s$ bands, the tilt is stronger, similarly to graphene \cite{BeugelingEA2012PRB86-07}. Here, the spin vector at the edge site points $\sim 30^{\circ}$ away from the vertical. This difference in tilt can be explained by the much larger intrinsic SOC in the $p$ sector than in the $s$ one, whereas the two Rashba couplings $\gamma_{ss\sigma}$ and $\gamma_{pp\sigma}$ are of similar order of magnitude. Strictly speaking, one should denote this state a $Z_2$ topological insulator, but since the Rashba SOC is extremely small, one can think of an approximate QSHE. 

{\bf Flat band and Coulomb interactions.} The large gaps in the $p$ sector are mainly due to the intrinsic SOC, which, contrarily to the Kane-Mele model, is described by an on-site term ($\lambda_\mathrm{ISO}^{p}=15.8\meV$). In particular, the intrinsic SOC separates the lowest flat band from the other $p$ bands, with large gaps, e.g., $35\meV$ in the case of Fig.~\ref{fig_bulkdispTB}c. This gap ranges from 13 meV to 36 meV when we vary the nanocrystal size and shape (Supplementary Table 1). 


\begin{table}
  \center%
  \begin{tabular}{cc|cc|c|cc}
    $\varepsilon_\mathrm{in}$ & $\varepsilon_\mathrm{out}$ & $U$ & $V$ & Band width & Absolute gap &  Gap at $\Gamma$ \\
    \hline
    $14$ & $14$ & $48\meV$ & $23\meV$ & \multirow{2}{*}{$20 \meV$} & \multirow{2}{*}{$35\meV$} & \multirow{2}{*}{$56 \meV$} \\
    $14$ & $6$  & $76\meV$ & $43\meV$ & &
  \end{tabular}
  \caption{{\bf Coulomb energies.} On-site ($U$) and nearest-neighbor ($V$) Coulomb interaction energies calculated for a honeycomb lattice of HgTe nanocrystals (with lattice parameters $q=0.5$, $a=5\nm$) compared to the width of the lowest $p$ band, the absolute gap between the lowest $p$ bands (the gap between band extrema), and the vertical gap at $\Gamma$ ($\vec{k}=0$) between them. $\varepsilon_\mathrm{in}$ and $\varepsilon_\mathrm{out}$ are the dielectric constants of the materials composing the lattice (HgTe) and its environment, respectively.}
  \label{tbl_coulomb}
\end{table}



\begin{figure*}[t]
\centering
\includegraphics[width=0.9\textwidth]{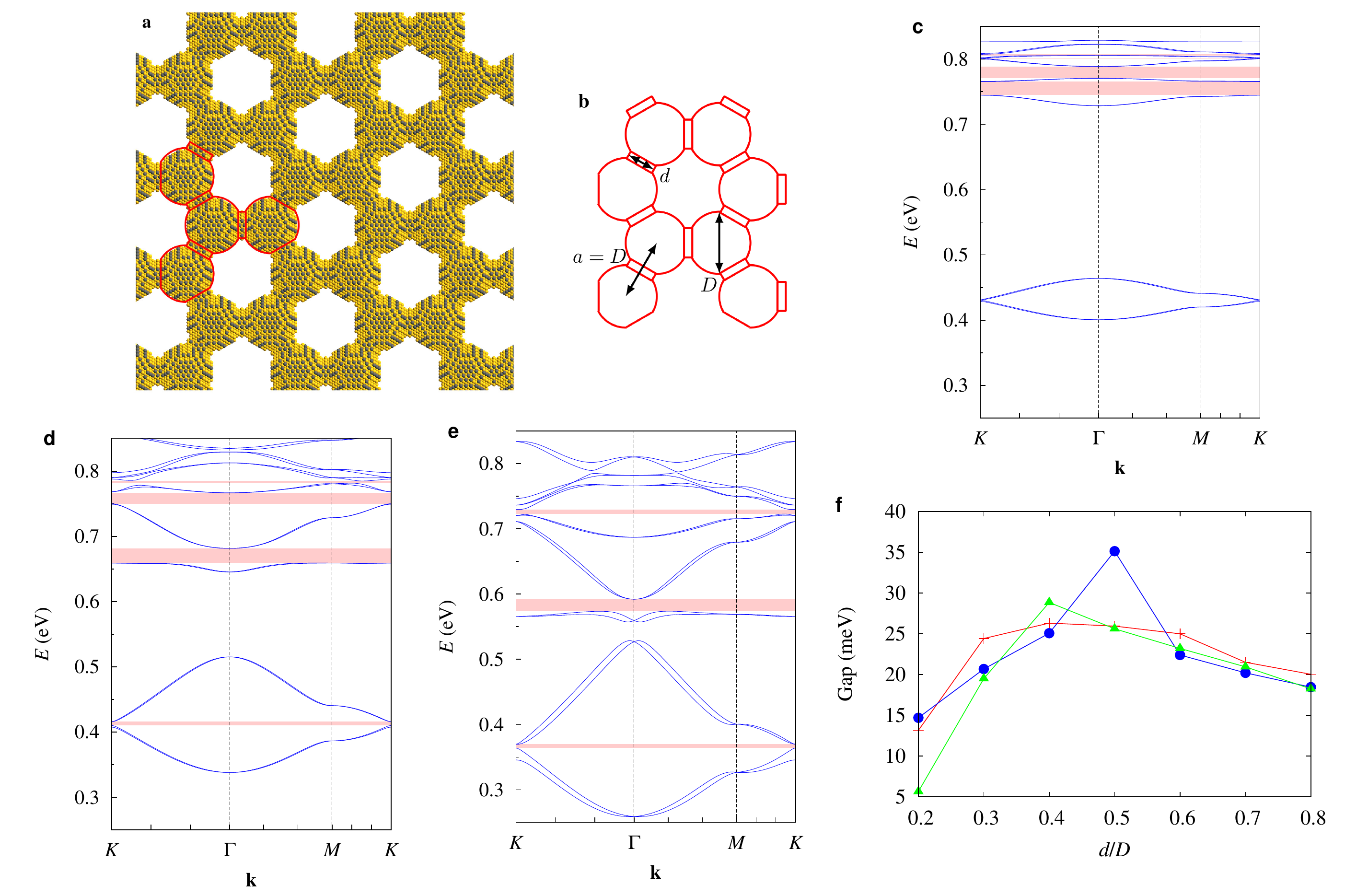}
\caption{\textbf{Honeycomb lattice of HgTe spheres and its conduction band structure.} \textbf{a,b}, top view of an assembly of spheres connected by cylinders, forming a honeycomb lattice of HgTe. Hg atoms are in yellow, Te atoms are in grey. The lattice spacing $a$, i.e., the center-to-center distance between neighbor spheres, is equal to the diameter $D$ of the spheres and $d$ is the diameter of the cylinders. \textbf{c,d,e}, conduction band dispersions resulting from the atomistic tight-binding calculation for $D=5.0\nm$ and $d/D=0.3$ (\textbf{c}), $d/D=0.6$ (\textbf{d}), or $d/D=0.8$ (\textbf{e}). Non-trivial gaps are indicated by pink shaded regions. \textbf{f}, evolution of the gap between the lowest $p$ bands versus $d/D$ for three values of $D$ (green triangles: $D=5.9 \nm$; blue circles: $D=5.0 \nm$; red crosses: $D=4.1 \nm$).}
  \label{fig_sph}
\end{figure*}


Under partial filling of the lowest-energy flat $p$ band by electrons, fascinating phenomena such as fractional QSHE are predicted in presence of strong correlations, when the strength of Coulomb interactions between electrons is large compared to the band width and smaller or comparable to the energy gap between the flat band and the next higher one (for a recent review, see Ref.~\onlinecite{Bergholtz13}). In addition, it has been shown that also for large Coulomb interactions, a fractional Chern insulator phase cannot be ruled out \cite{Kourtis14}. We have thus calculated the on-site ($U$) and NN ($V$) electron-electron interaction energies for the honeycomb lattice of Fig.~\ref{fig_bulkdispTB}a, assuming two different dielectric environments around the nanocrystals (Methods). Table~\ref{tbl_coulomb} shows that Coulomb energies are larger than the band width and are comparable to the gap between $p$ bands. Therefore, the gap sizes open the possibility to the experimental observation of strongly correlated quantum phases including the long sought fractional QSHE\cite{NeupertEA2011,SunEA2011,HuEA2011,RegnaultBernevig2011,Goerbig2012}. 


{\bf Band structure for other types of HgTe lattices.} In order to understand the effects of the electronic coupling between honeycomb lattice sites, we have studied a second type of superlattices with a simpler geometry, consisting of tangent spheres connected by horizontal cylinders (Fig.~\ref{fig_sph}a,b). The HgTe spheres have their $[111]$ axis orthogonal to the lattice plane and the cylinders are oriented along $\langle 110 \rangle$ directions perpendicular to $[111]$.
Figures~\ref{fig_sph}c,d,e depict the evolution of the band structure with the ratio between the diameters of cylinders ($d$) and spheres ($D$) (see also Supplementary Fig.~3). An increase of $d/D$ induces larger NN hopping terms, broader $s$ and $p$ bands, and stronger $sp$ hybridization, as shown by the deformation of the $s$ band. It also results in a larger NNN term $\lambda_\mathrm{ISO}^{s}$, explaining why the non-trivial gap in the $s$ sector only exists for $d/D > 0.3$. On the contrary, topological gaps are always present in the $p$ sector, even for small values of $d/D$, because they are mainly determined by the on-site term $\lambda_\mathrm{ISO}^{p}$. At high values of $d/D$ ($>0.7$), the spin splitting of the bands in the entire Brillouin zone becomes particularly important due to increased Rashba couplings. In general, the lowest $p$ band is rather flat and has a maximum separation from the next higher one for $d/D$ close to $0.4$--$0.5$. In that case, the energy separation can be as large as $35\meV$ (Fig.~\ref{fig_sph}f). HgTe honeycomb structures with moderate coupling between lattice sites should provide the most suitable gap and band widths to observe the strongly correlated phases associated to the flatness of the bands.

In Supplementary Fig.~4, we show further results for a third type of honeycomb structure made of overlapping HgTe cylinders parallel to each other. Once again, the band structure can be described by the effective model (Supplementary Note 3 and Supplementary Fig.~5). In that case, the NN couplings are even stronger, there is no gap in the $s$ sector due to a large Rashba term, but the non-trivial gap above the lowest $p$ band remains. We can conclude that the topological effects in the $p$ sector are robust against changes in the electronic coupling and honeycomb period.

\hfill \break

{\bf Discussion}

In summary, we have performed atomistic TB calculations of the band structure of 2D honeycomb lattices of HgTe. We demonstrate that the strong SOC of HgTe combined with the honeycomb structure results in several topological phases. The calculated band structure can be described by a honeycomb lattice model as in graphene but including multi-orbital degrees of freedom that generate in particular a topologically non-trivial flat band. By taking advantage of these features, we show that in the same structure not only the QSHE, but potentially also the elusive fractional QSHE could be observed, just by varying the electron density. Both topological effects turn out to be protected by a gap as large as $35\meV$, and can thus be observed at high temperatures. Honeycomb superlattices of HgTe are therefore platforms of high interest to study electrons on a multi-orbital honeycomb lattice under strong SOC. Such structures could be fabricated by nanocrystal self-assembly (in a similar way as for PbSe and CdSe \cite{EversEA2013,BoneschanscherEA2014}) or by a combination of gas-phase deposition and lithography. Our results open the path towards high-temperature quantum spintronics in artificial graphene. 

\hfill \break

{\bf Methods}

\footnotesize%
{\bf Atomistic TB methodology.} The electronic structure of HgTe superlattices is calculated within the TB approximation, as described in detail in Ref.~\onlinecite{Allan12}. The TB Hamiltonian matrix is written in a basis of atomic orbitals ($sp^3d^5s^*$ for each spin orientation) as function of parameters that have been obtained by fitting to two reference band structures: Close to the Fermi level, we use the $\vec{k} \cdot \vec{p}$ band structure of Ref.~\onlinecite{Man91} whereas elsewhere we use the band structure of Ref.~\onlinecite{Svane11} obtained using a quasi-particle self-consistent GW approximation in a hybrid scheme. In the present work, we have used the TB parameters that give the band structure of HgTe at $300\,\mathrm{K}$. The surfaces of the superlattices are saturated by pseudo-hydrogen atoms that push surface states far from the energy regions of interest in this study. Therefore surface states do not interfere with edge states predicted in ribbons. Due to the large size of the systems that we have studied (up to $\approx 10^5$ atoms per unit cell), only near-gap eigenstates are calculated using the numerical methods described in Ref.~\onlinecite{Niquet00}.

{\bf Calculation of the $Z_2$ topological invariant.} The $Z_2$ topological invariant (spin Chern number $C_s$) for the bands of interest is calculated using the model Hamiltonian following the methodology proposed in Ref.~\onlinecite{Fukui07} and derived from Ref.~\onlinecite{Fu06}. This approach works even for systems without inversion symmetry, which is the case here. $C_s$ is given by a sum of terms calculated on a regular lattice in the Brillouin zone. We have checked that the results converge for a mesh denser than $21 \times 21$ $k$ vectors. In all cases, the invariants that we have computed for the bands are consistent with the number of edge states we observe in the bulk gaps.

{\bf Coulomb interactions.} The Coulomb repulsion between electrons in honeycomb lattices of HgTe nanocrystals can be characterized as follows. For simplicity, we consider electrons in the $s$ band, since in the case of individual nanocrystals, it was shown theoretically \cite{Niquet02} and experimentally \cite{Banin99} that the Coulomb integrals are almost identical for states with $s$ and $p$ envelope functions. The Coulomb interaction associated to electrons on nanocrystals $i$ and $j$ is calculated as
\begin{equation}
\int  |\Psi_{i}(\vec{r})|^{2} |\Psi_{j}(\vec{r}')|^{2} {\cal V}(\vec{r},\vec{r}') d\vec{r} d\vec{r}'
\end{equation}
where $\Psi_{i}(\vec{r})$ is the $s$ state on the nanocrystal $i$, and ${\cal V}(\vec{r},\vec{r}')$ is the screened Coulomb energy of two interacting electrons at $\vec{r}$ and $\vec{r}'$. Taking into account that there is just one (spin-degenerate) $s$ state per nanocrystal, $\Psi_{i}$ and $\Psi_{j}$ are simply defined as the components on nanocrystals $i$ and $j$ of the wavefunction calculated for the lowest $s$ band at $\Gamma$ (or another $k$ vector), normalized on their respective nanocrystals. Coulomb matrix elements are decomposed in the basis of the atomic orbitals and are calculated following the methodology described in Ref.~\onlinecite{Delerue97}, by using usual approximations in the TB description, i.e., neglecting overlaps between atomic orbitals and considering atomic charges as point-like charges. For the potential ${\cal V}$, we consider two configurations: First, a dielectrically homogeneous system, for which ${\cal V}(\vec{r},\vec{r}') = e^{2}/ \left ( \varepsilon_{\mathrm{in}} |\vec{r} - \vec{r}'| \right )$, where $\varepsilon_{\mathrm{in}}=14$ is the dielectric constant of HgTe; Second, a dielectrically inhomogeneous system, for which ${\cal V}$ is calculated by solving the Poisson equation,  with the dielectric constant inside (outside) the lattice equal to $\varepsilon_{\mathrm{in}}$ ($\varepsilon_{\mathrm{out}}$). We have chosen $\varepsilon_{\mathrm{out}}=6$, a typical value taken to simulate the complex dielectric environment around semiconductor nanocrystals \cite{Niquet02}. The on-site ($U$) and NN ($V$) terms are presented in Table~\ref{tbl_coulomb}. As expected, larger values are obtained for $\varepsilon_{\mathrm{out}}=6$ than for $\varepsilon_{\mathrm{out}}=\varepsilon_{\mathrm{in}}$.  Longer-range Coulomb terms are expected to decay as the inverse of the distance between nanocrystals. However, it is important to note that these values do not take into account the extra screening induced by the electrons filling the bands. This could be computed, for example, in the random-phase approximation, but this is clearly beyond the scope of the present work. Long-range interactions will be strongly screened, while short-range ones will be only slightly reduced \cite{Nozieres99}. In this context, for a band filling of the order of $1/3$, correlations will be mainly governed by short-range effects.

\hfill \break
\normalsize%
{\bf Acknowledgements}

This work was supported by the French National Research Agency (ANR) project "ETSFG" (ANR-09-BLAN-0421-01). W.B. is funded by the Max-Planck-Gesellschaft through the visitor's programme at MPI-PKS. E.K. acknowledges funding by the University of Luxembourg Research Office. D.V. and C.M.S. wish to acknowledge the Dutch FOM for financial support via the Program 13DDC01 "Designing Dirac Carriers in Honeycomb Superlattices". The work of C.M.S. is part of the D-ITP consortium, a program of the Netherlands Organisation for Scientific Research (NWO) that is funded by the Dutch Ministry of Education, Culture and Science (OCW).

\hfill \break
{\bf Author contributions}

W.B., E.K., and C.D. performed the calculations. Y.M.N. contributed to the development of the codes and methodologies. C.D., D.V. and C.M.S. supervised the project. All authors were involved in writing of the manuscript.

\hfill \break
{\bf Additional information}

{\bf Competing financial interests:} The authors declare no competing financial interests.

\medskip{\bf How to cite this article:} Beugeling, W.~\emph{et al.}, Topological states in multi-orbital HgTe honeycomb lattices. \emph{Nat.~\mbox{Commun.}} 6:6316 \href{http://dx.doi.org/10.1038/ncomms7316}{doi:10.1038/ncomms7316} (2015).

\medskip\footnotesize%
This work is licensed under a Creative Commons Attribution 4.0
International License. The images or other third party material in this
article are included in the article’s Creative Commons license, unless indicated otherwise
in the credit line; if the material is not included under the Creative Commons license,
users will need to obtain permission from the license holder to reproduce the material.
To view a copy of this license, visit \href{http://creativecommons.org/licenses/by/4.0/}{http://creativecommons.org/licenses/by/4.0/}

\normalsize

\vfill\clearpage

\onecolumngrid

\begin{figure*}[h]
\includegraphics[width=0.33\textwidth]{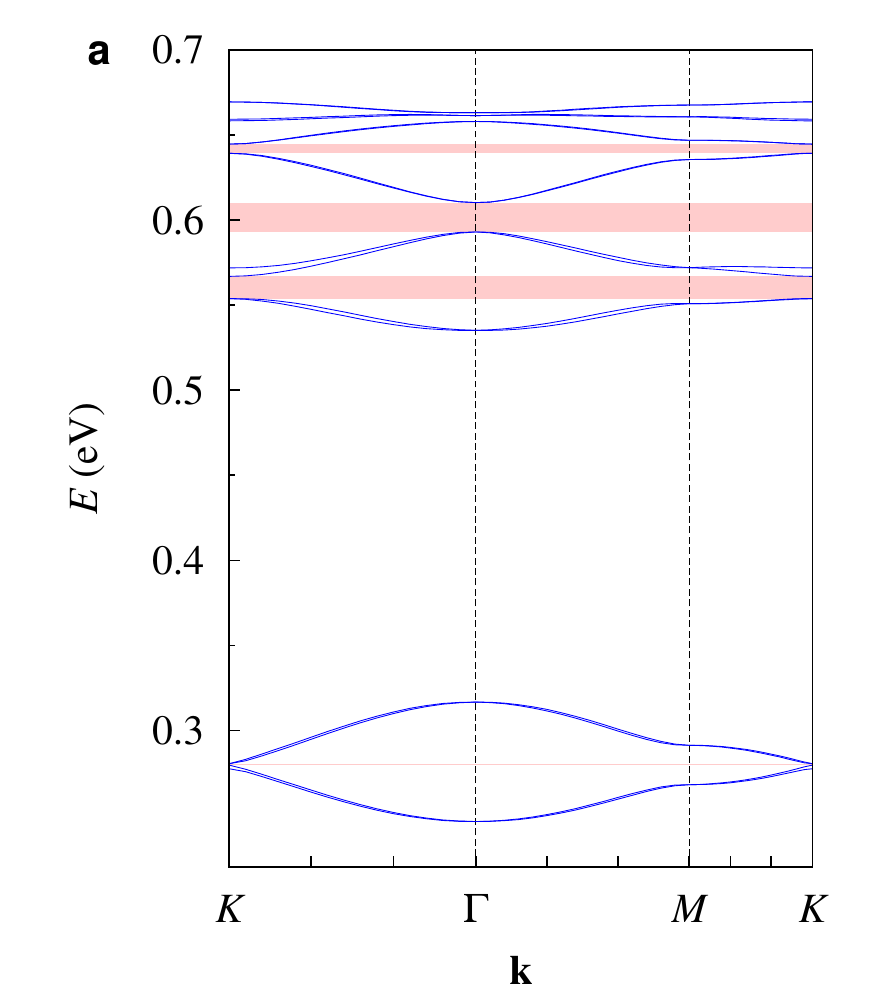}%
\includegraphics[width=0.33\textwidth]{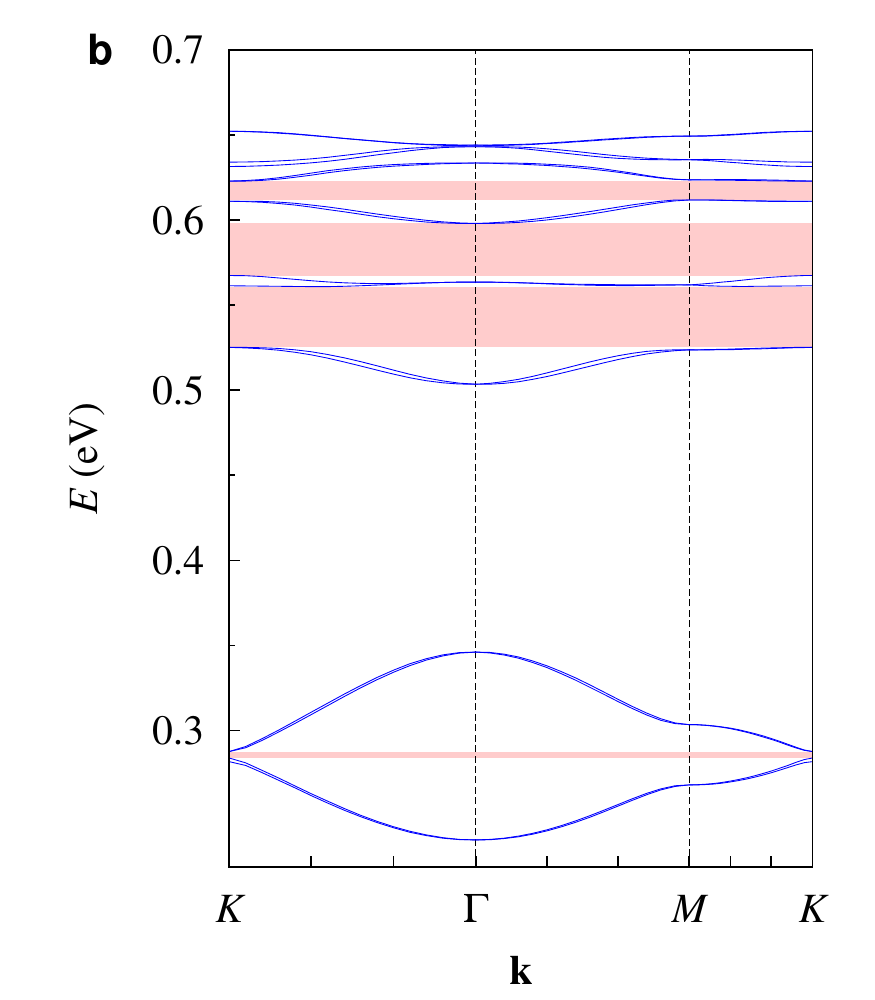}\\
\includegraphics[width=0.33\textwidth]{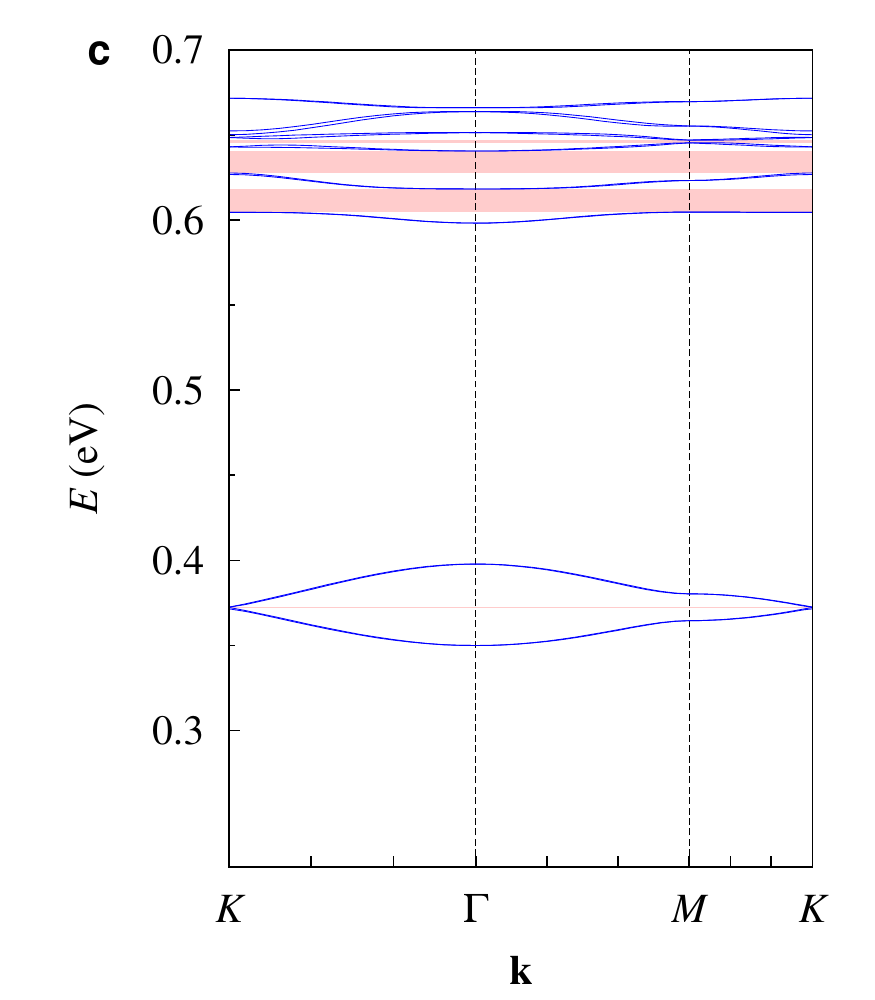}%
\includegraphics[width=0.33\textwidth]{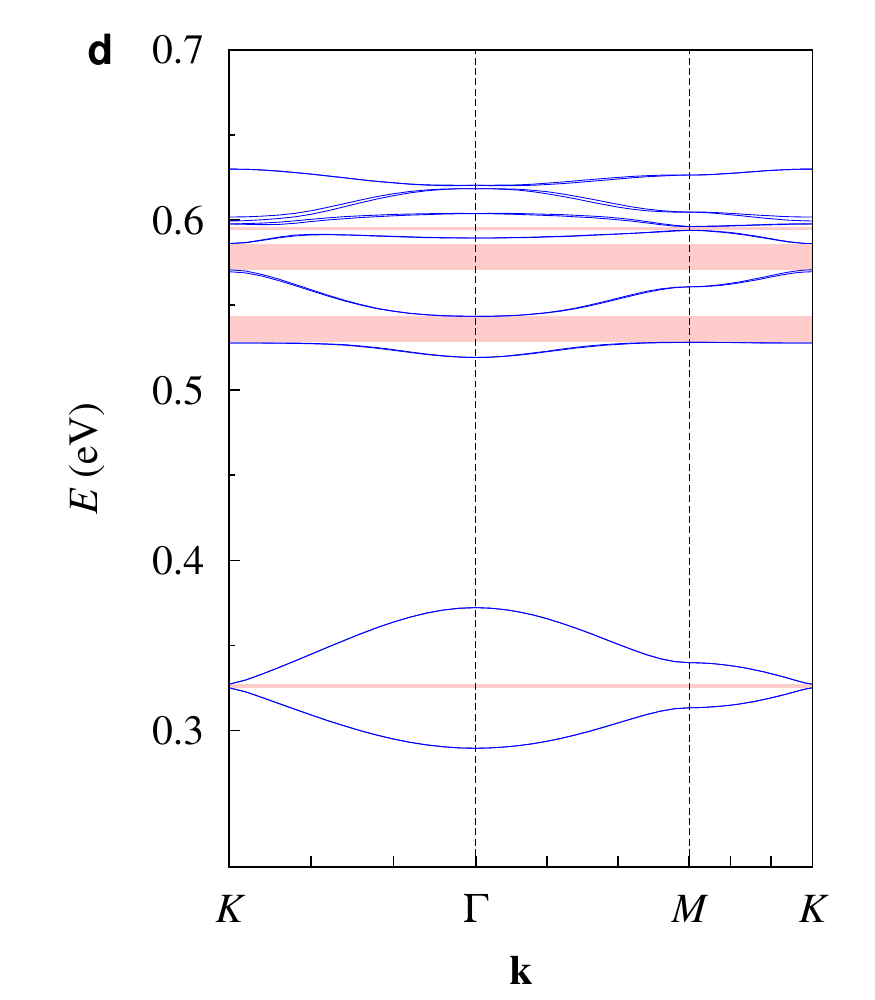}\\[1em]
\suppcaption{{\bf Supplementary Figure 1: Conduction bands in lattices of HgTe nanocrystals.} Band
dispersion resulting from the atomistic tight-binding calculations. Topologically non-trivial
gaps are indicated by pink shaded regions. Each nanocrystal has a truncated nanocube shape.
Truncation factor: $q=0.25$ (\textbf{a,c}) or $q=0.45$ (\textbf{b,d}). Honeycomb lattice spacing: $a=5.9\nm$ (\textbf{a,b}) or $a=6.8\nm$ (\textbf{c,d}).}
\end{figure*}
\clearpage

\begin{figure*}[!h]
\includegraphics[height=60mm]{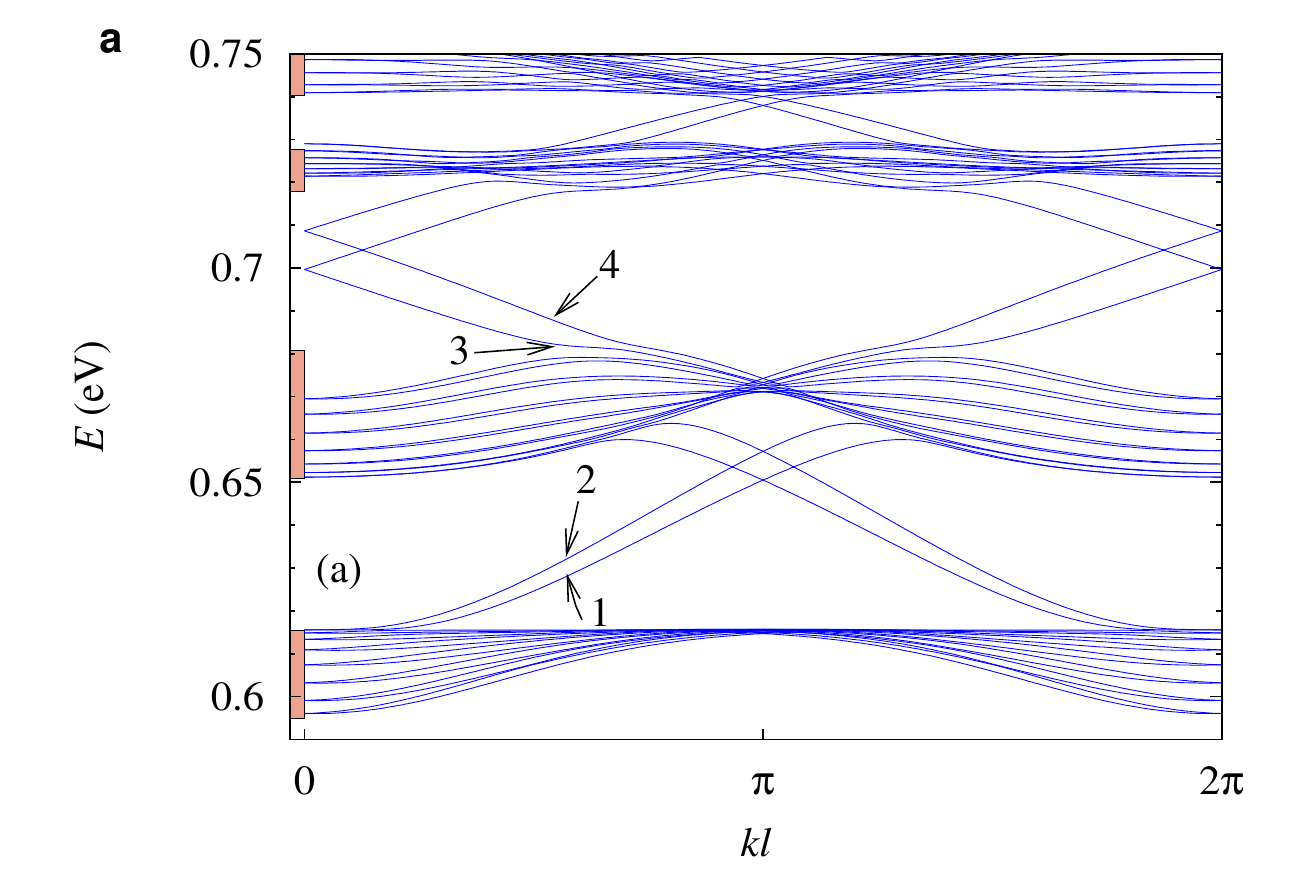}%
\raisebox{8mm}{\includegraphics[height=50mm]{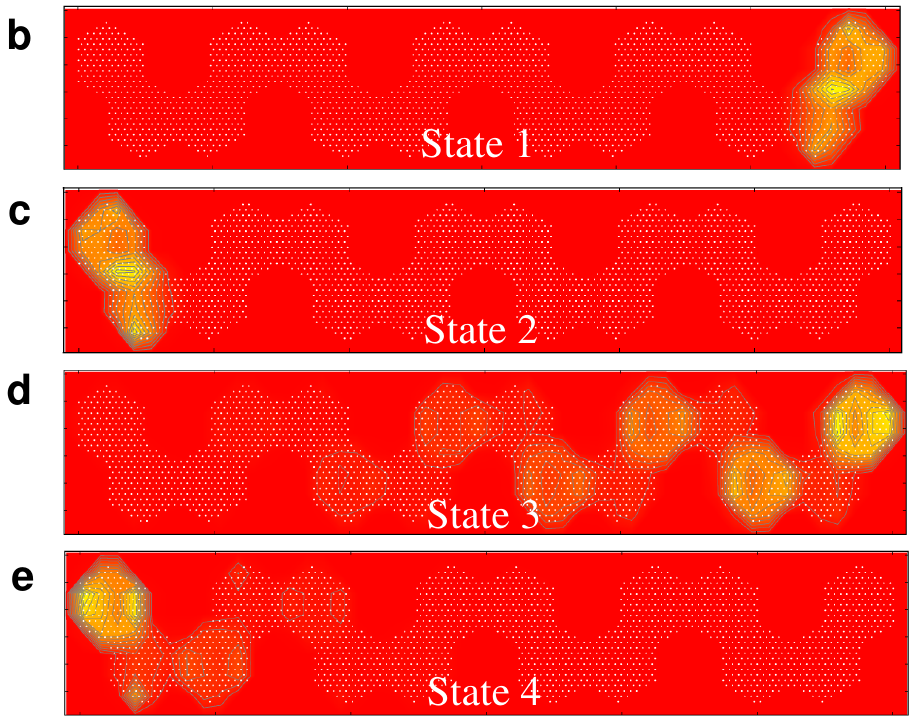}}\\
\includegraphics[width=70mm]{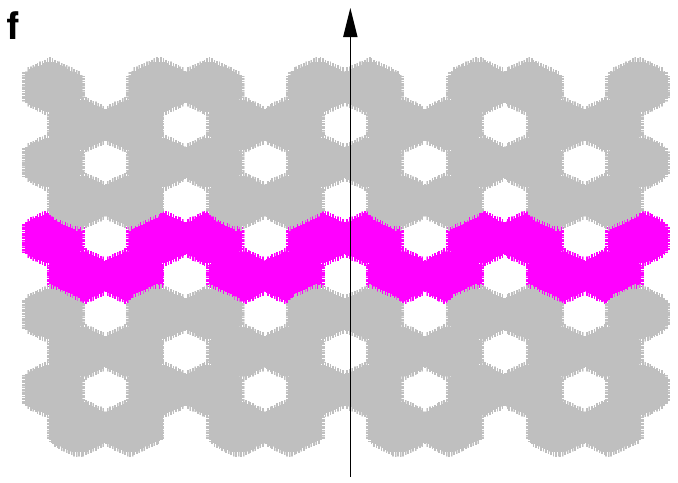}\hspace{50mm}\phantom{.}\\[1em]
 \suppcaption{{\bf Supplementary Figure 2: Ribbon with broken inversion symmetry.} Atomistic tight-binding
calculations for a honeycomb ribbon of HgTe nanocrystals, with broken inversion symmetry.
The nanocrystals have a truncated nanocube shape (truncation factor $q=0.5$, honeycomb lattice
spacing $a=5.0\nm$). A single plane of atoms is removed from the right side of the ribbon in
order to break the inversion symmetry (not shown). \textbf{f}, schematic view of the ribbon. The unit
cell of 16 nanocrystals, shown in magenta, is reproduced periodically along the direction
indicated by the arrow. \textbf{a}, dispersion of the p bands. The position of the bulk bands is indicated
by pink vertical bars along the left axis. \textbf{b,c,d,e}, 2D plots of the wave functions of four states
calculated at $k = 0.3\times2\pi/l$ where $l$ is the length of the unit cell. The labels 1 (\textbf{b}), 2 (\textbf{c}), 3 (\textbf{d})
and 4 (\textbf{e}) refer to the states indicated in \textbf{a}. The plots are restricted to the unit cell of the ribbon.
The white dots indicate the atoms.}
\end{figure*}
\clearpage

\begin{figure*}[!h]
\includegraphics[width=0.33\textwidth]{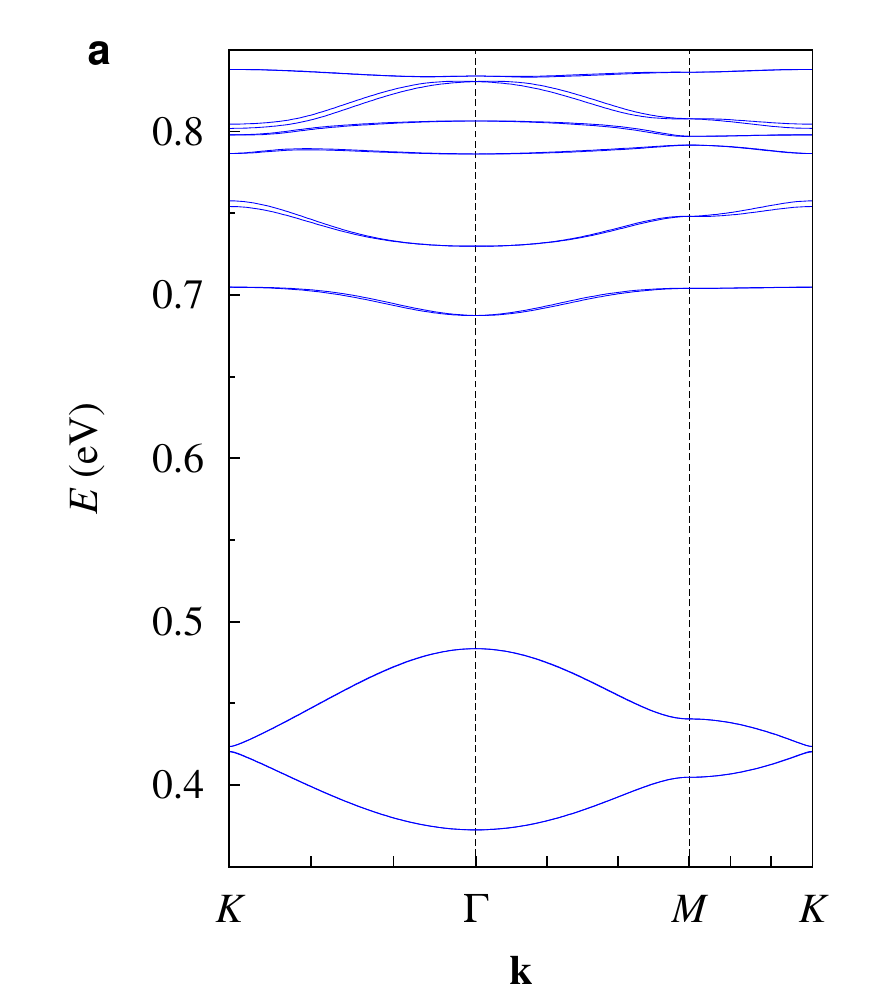}%
\includegraphics[width=0.33\textwidth]{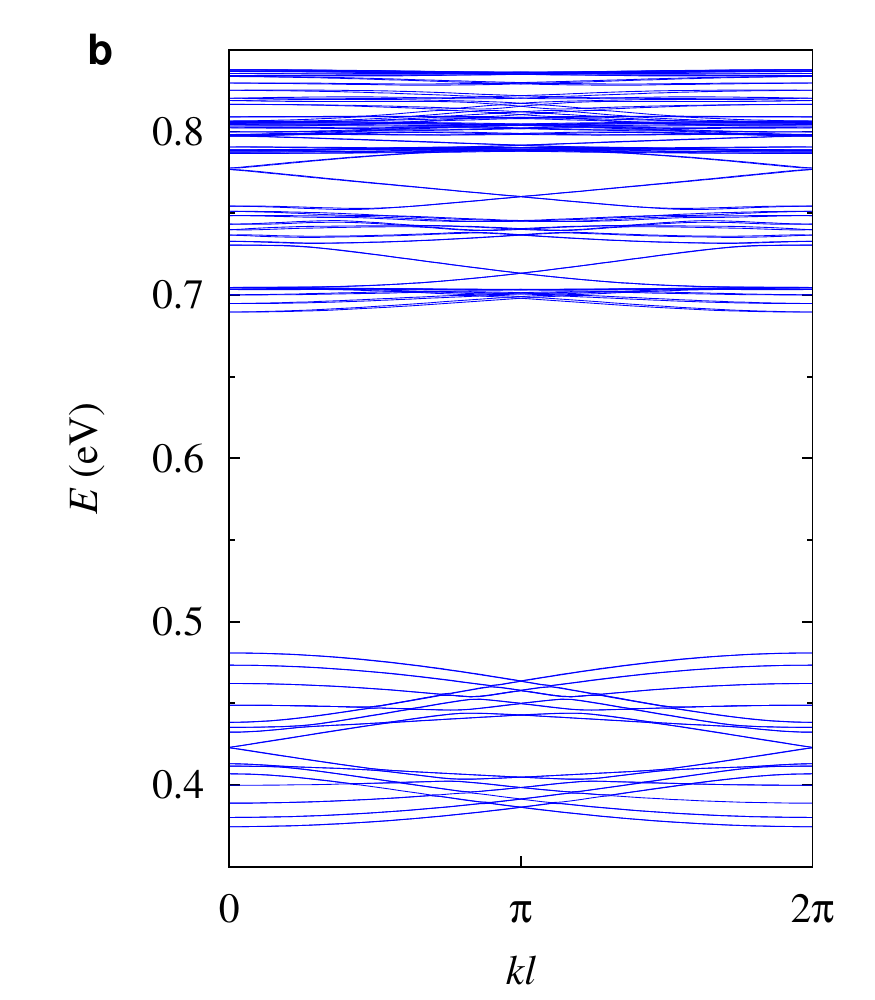}\\[1em]
 \suppcaption{{\bf Supplementary Figure 3: Band structure for lattices of spherical nanocrystals.} Conduction
band dispersion resulting from the atomistic tight-binding calculation for honeycomb
superlattices of spheres (diameter $D=5\nm$) connected by cylinders (diameter $d=0.4 D$). \textbf{a}: bulk
(gap between the lowest $p$ bands $= 25\meV$). \textbf{b}: armchair nanoribbon composed of sixteen
nanocrystals per unit cell.
}
\end{figure*}

\begin{figure*}[!h]
\includegraphics[width=50mm]{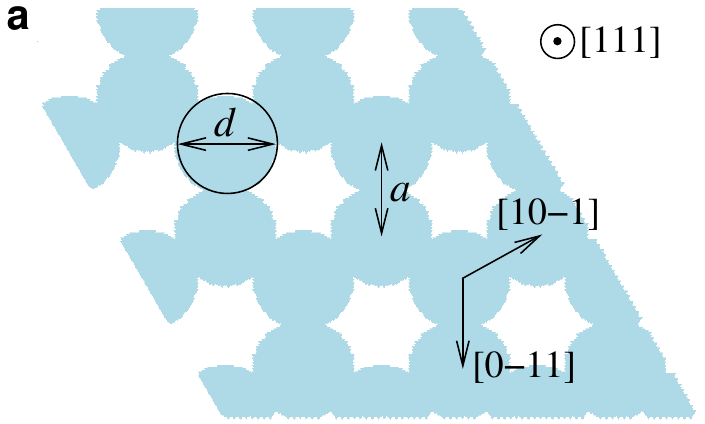}\hspace{55mm}\phantom{.}\\
\includegraphics[width=0.33\textwidth]{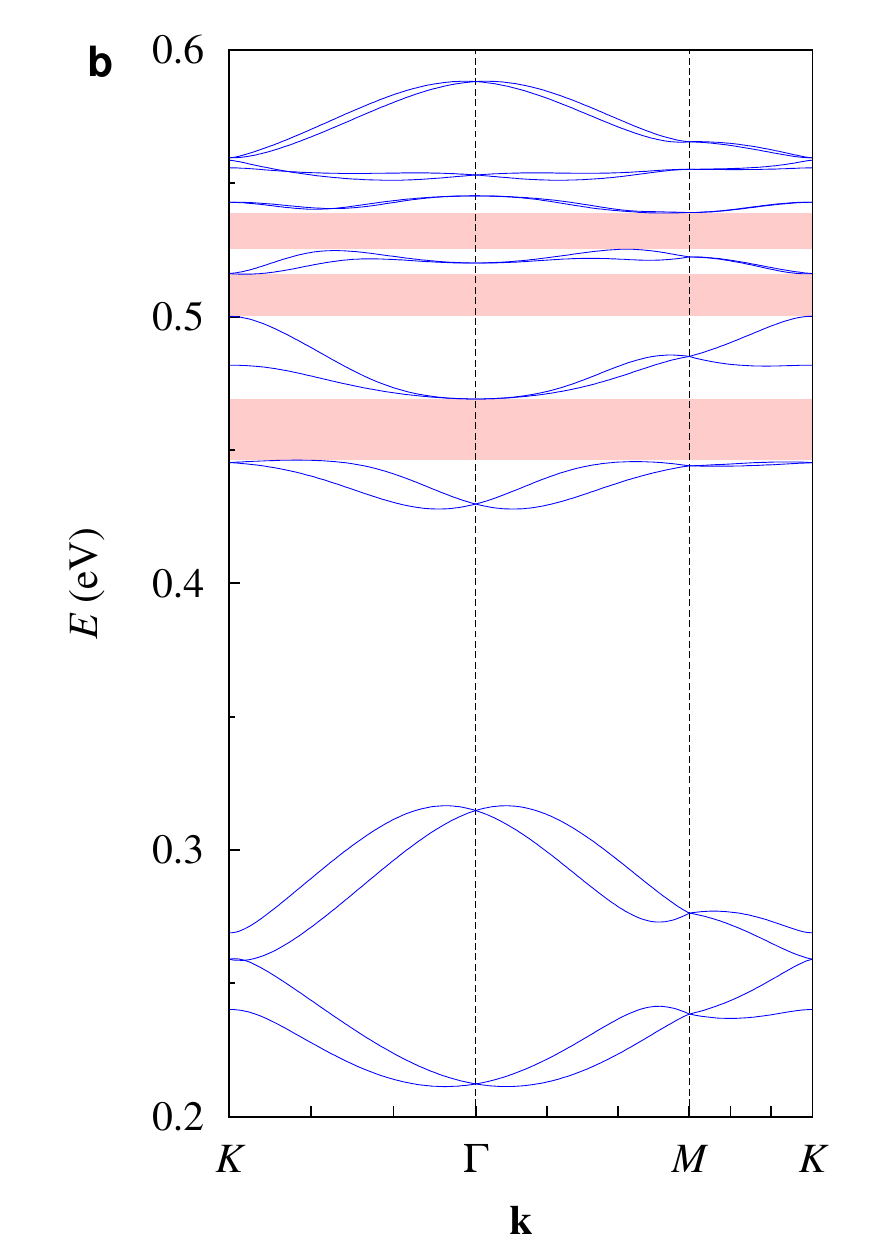}%
\includegraphics[width=0.33\textwidth]{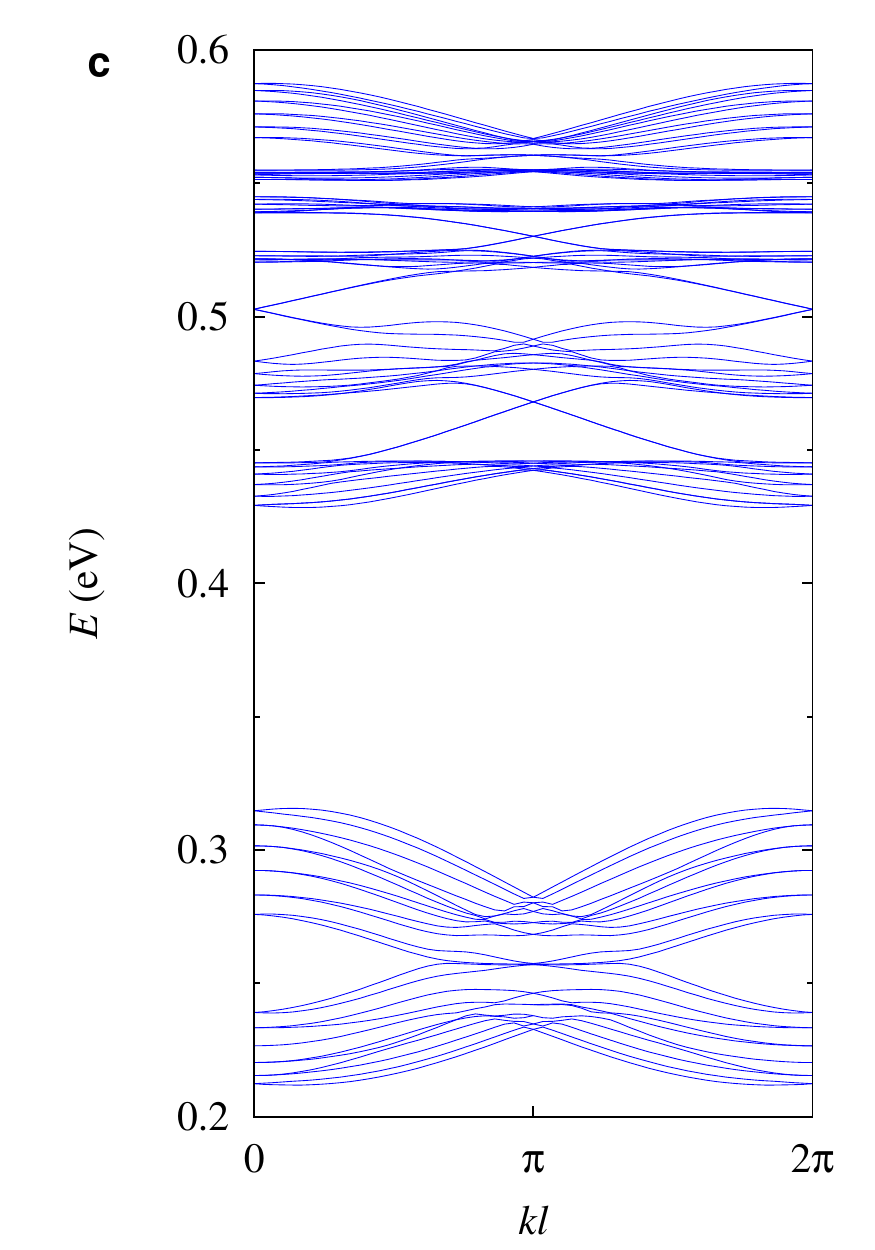}\\[1em]
 \suppcaption{{\bf Supplementary Figure 4: Band structures for lattices of cylinders.} \textbf{a}, top view of an
assembly of vertical cylinders forming a honeycomb lattice of HgTe (cylinder diameter $d =
1.02a = 7.0\nm$, layer thickness $t = 5.3\nm$, $a$ is the lattice spacing). Crystallographic axes of
HgTe are indicated. \textbf{b}, corresponding conduction band dispersion resulting from the atomistic
tight-binding calculation. Non-trivial gaps are indicated by pink shaded regions. \textbf{c}, band
structure of a zigzag ribbon formed of $12$ cylinders per unit cell.
}
\end{figure*}
\clearpage

\begin{figure*}[!h]
\includegraphics[width=0.33\textwidth]{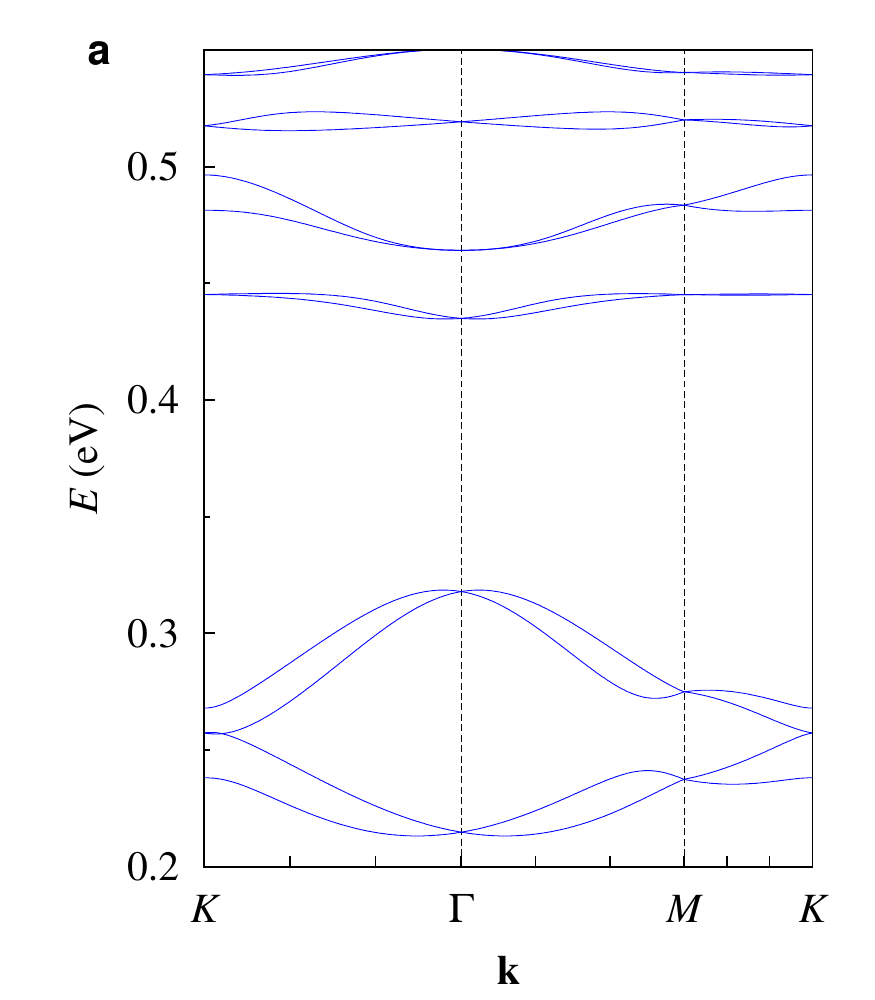}%
\includegraphics[width=0.33\textwidth]{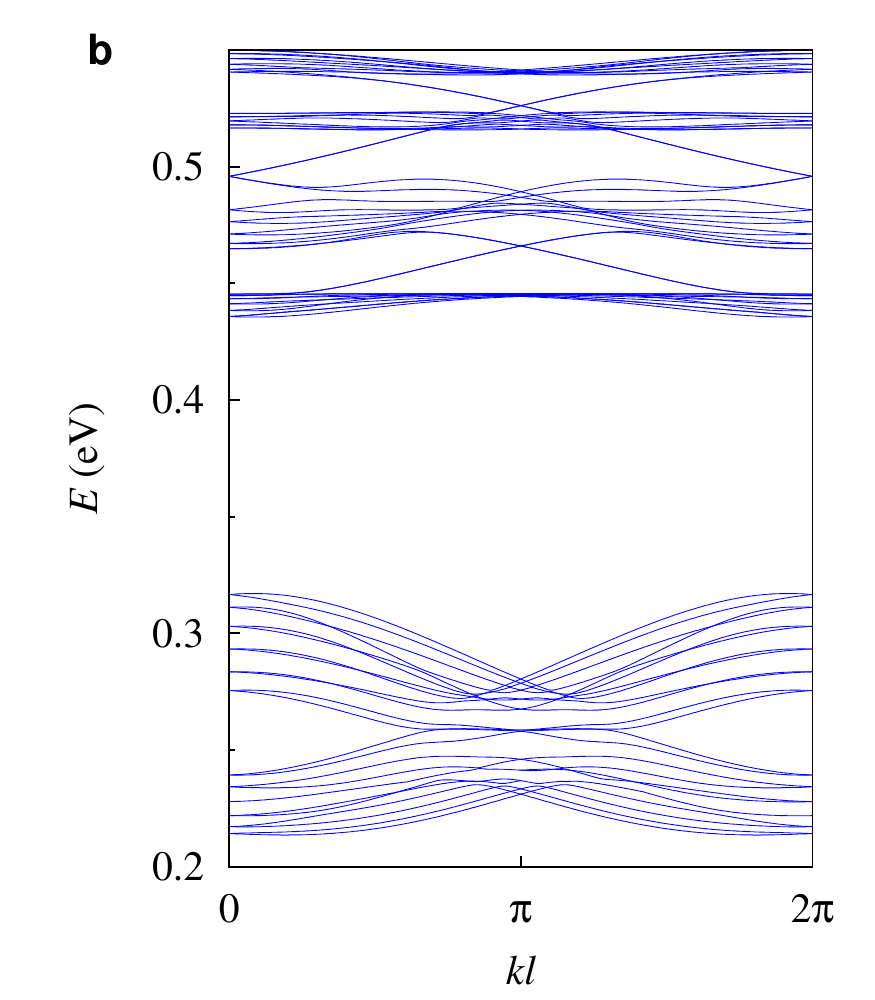}\\[1em]
 \suppcaption{{\bf Supplementary Figure 5: Effective model for a lattice of cylinders.} \textbf{a}, conduction band
dispersion calculated using the model Hamiltonian. \textbf{b}, band structure of a zigzag ribbon formed
by $12$ sites per unit cell.}

\end{figure*}

\begin{table*}[!h]
 \begin{tabular}{c|ccc}
  & $a = 5.0\nm$&$a = 5.9\nm$&$a = 6.8\nm$\\
  \hline
  $q = 0.25$&$17.2$&$13.1$&$13.7$\\
  $q = 0.30$&$17.2$&$13.6$&$13.7$\\
  $q = 0.35$&$15.3$&$13.6$&$32.0$\\
  $q = 0.40$&$15.3$&$36.2$&$32.0$\\
  $q = 0.45$&$35.4$&$36.2$&$15.7$\\
  $q = 0.50$&$35.4$&$18.6$&$15.7$
 \end{tabular}
\\[1em]
\suppcaption{{\bf Supplementary Table 1: Energy gap in the $p$ bands.} Energy gap ($\mathrm{meV}$) between the lowest $p$
bands in honeycomb lattices of HgTe nanocrystals for different values of the truncation factor $q$
and the honeycomb lattice spacing $a$. The same gap predicted for two successive values of $q$
means that the honeycomb lattices are identical, the number of atoms per nanocrystal varying
discontinuously with $q$.}
\end{table*}

\begin{table*}[!h]
 \begin{tabular}{llll}
  on-site&NN hopping&Rashba SOC&intrinsic SOC\\
  \hline
  $E_s=266$  & $V_{ss\sigma}=-17.2$ & $\gamma_{ss\sigma}=5.11$ & $\lambda^s_\mathrm{ISO}=0.3$\\
  $E_{p_x}=493$ & $V_{pp\sigma}=28.9$ & $\gamma_{pp\sigma}=4.77$ & $\lambda^p_\mathrm{ISO}=14.1$\\
  $E_{p_y}=493$ & $V_{pp\pi}=-0.6$ & $\gamma_{pp\pi}=0.00$\\
  $E_{p_z}=698$ & $V_{sp\sigma}=24.2$\\
 \end{tabular}
\\[1em]
\suppcaption{{\bf Supplementary Table 2: Parameters of the effective model.} Parameters ($\mathrm{meV}$) used for the
model for the HgTe superlattice described in Supplementary Fig.~5a. $E_s$, $E_{p_x}$, $E_{p_y}$, and $E_{p_z}$ are
the onsite energies on the $s$, $p_x$, $p_y$, and $p_z$ orbitals, respectively. $V_{ss\sigma}$, $V_{pp\sigma}$, $V_{pp\pi}$, and $V_{sp\sigma}$ are the
hopping parameters, following the notations of Ref. [1]. $\gamma_{ss\sigma}$ , $\gamma_{pp\sigma}$, and $\gamma_{pp\pi}$ are the terms describing the Rashba SOC, following the same notations. The intrinsic SOC is defined by $\lambda^s_\mathrm{ISO}$
and $\lambda^p_\mathrm{ISO}$ on $s$ and $p$ orbitals, respectively.}
\end{table*}

\bigskip
{\center\bf Supplementary Note 1\\
Influence of size on lattices of HgTe nanocrystals.\\}

\medskip When we vary the lattice spacing $a$ ($3$--$8\nm$) and the nanocube truncation ($q$ between $0.25$ and
$0.5$) in lattices of HgTe nanocrystals, all band structures have similar behaviour. Examples of
band structure are shown in Supplementary Fig.~1. The presence of well-separated $s$ and $p$
bands with helical gaps is quite general, even if the width of the gaps may vary substantially
depending on the geometry of the superlattices. The overall shape of the $s$ bands is always the
same, whereas for the $p$ bands the variations are more important because not only the nearest-
neighbour hopping, but also the respective positions of the $p_x$, $p_y$ and $p_z$ states depend on
nanocrystal size and truncation. However, the lowest $p$ band is always detached from the next
higher one. The width of the lowest gap in the $p$ bands is given in Supplementary Table 1 for
different values of $a$ and $q$.

\bigskip
{\center\bf Supplementary Note 2\\
Edge localization in ribbons of HgTe nanocrystal superlattices\\}

\medskip In this section, we present additional results on ribbons made from lattices of HgTe
nanocrystals. We consider the same nanocrystals as in Figs. 1 and 2 ($q=0.5$, $a=5.0\nm$) but we
investigate a ribbon in which the two edges are not symmetric by inversion. First, we discuss
the effect of this asymmetry on the band structure. Second, we present plots of the wave
functions.
When the two edges of the ribbon are equivalent by inversion symmetry, the topological edge
states on the opposite sides of the ribbon are quasi-degenerate for each value of $k$, the coupling
between opposite edge states being negligible (Fig.~2). In order to observe the effect of
geometry on the results, we have also considered a ribbon in which the inversion symmetry has
been broken. For that purpose, we have removed all atoms of the last atomic plane on the right
side of the ribbon and we have saturated the broken bonds with pseudo-hydrogen atoms.
Supplementary Fig.~2a shows that the edge states are preserved thanks to their topological
protection but their degeneracy at a given $k$ is lifted due to the asymmetry between the two
sides of the ribbon.
The 2D plots of the wavefunctions of the four states denoted 1--4 in Supplementary Fig.~2a for
the asymmetric ribbon are shown in Supplementary Figs.~2b--e. These states calculated at
$k = 0.3\times 2\pi/l$ are strongly localized on the edges of the ribbon. State 3 is more delocalized
than the other three states because it lies very close to the bulk band edge.

\bigskip
{\center\bf Supplementary Note 3\\
Band structures of honeycomb lattices of vertical cylinders\\}

\medskip We have investigated a third type of honeycomb structure composed of HgTe cylinders. The
axes of the cylinders are parallel to each other and are organized on a honeycomb lattice
(Supplementary Fig.~4a). Such structures could be fabricated from a HgTe layer, grown for
example by gas-phase approaches. The honeycomb nanogeometry is then defined using
nanoscale lithography.
Quite similar band structures are obtained for these lattices (Supplementary Fig.~4b). Once
again, $s$ and $p$-like bands can be easily identified. The Rashba SOC is in general much larger
than for lattices of nanocrystals due to a stronger coupling between neighbouring sites.
Interestingly, a very similar behaviour was found for superlattices of spheres connected by
cylinders when the coupling between neighbouring spheres is strong, i.e., for large values of
$d/D$ (Fig.~3). As a consequence of the large Rashba SOC, the gap in the $s$ sector is closed. Non-
trivial gaps remain in the $p$ bands, for many configurations that we have investigated, in spite of
larger spin splitting. However, the larger Rashba SOC tends to increase the dispersion of the
lowest $p$ band.
Once again, the results of the atomistic TB calculations are well described by the effective
model (Supplementary Table 2 and Supplementary Fig.~5a). Only the highest bands of
Supplementary Fig.~4b are not reproduced by the effective model because they involve higher-energy orbitals which are not considered in the model. The topological properties of the bands
are demonstrated by edge-state analyses in ribbons, using the atomistic TB calculations
(Supplementary Fig.~4c) or the effective TB model (Supplementary Fig.~5b) which give results
in excellent agreement. The non-trivial topology of the bands is confirmed by calculations of
the $Z_2$ topological invariants using the effective-model Hamiltonian.

\bigskip
{\center \bf Supplementary Reference\\}

\medskip\noindent [1] Slater, J. C. \& Koster, G. F.\quad Simplified LCAO method for the periodic potential problem.
\emph{Phys.\ Rev.} \textbf{94}, 1498--1524 (1954).

\end{document}